\documentclass[12pt,preprint]{aastex}

\renewcommand{\vec}[1]{{\mbox{\boldmath $#1$}}}

\newcommand{\etal}{et al. }

\newcommand{\eps}    {Earth Planets Space}
\newcommand{\icrc}[2]{Proc. #1 Int. Cosmic Ray Conf. (#2)}
\newcommand{\pr}     {Phys. Rev.}

\shorttitle{COSMIC RAY PENETRATION INTO MAGNETIC FLUX ROPE}
\shortauthors{Kubo and Shimazu}

\begin{document}

\title{Effect of Finite Larmor Radius on the Cosmic Ray Penetration into an Interplanetary Magnetic Flux Rope}

\author{Y\^uki Kubo\altaffilmark{1,3} AND Hironori Shimazu\altaffilmark{2,3}}
\affil{\altaffilmark{1}Space Environment Group, National Institute of Information and Communications Technology, Tokyo, 184-8795, JAPAN; kubo@nict.go.jp}
\affil{\altaffilmark{2}Knowledge Creating Communication Research Center, National Institute of Information and Communications Technology, Kyoto, 619-0289, JAPAN}
\affil{\altaffilmark{3}Department of Information System Fundamentals, The University of Electro-Communications, Tokyo, 182-8585, JAPAN}
%\email{kubo@nict.go.jp}

\begin{abstract}
We discuss a mechanism for cosmic ray penetration into an interplanetary magnetic flux rope, particularly the effect of the finite Larmor radius and magnetic field irregularities. First, we derive analytical solutions for cosmic ray behavior inside a magnetic flux rope, on the basis of the Newton-Lorentz equation of a particle, to investigate how cosmic rays penetrate magnetic flux ropes under an assumption of there being no scattering by small-scale magnetic field irregularities. The results show that the behavior of a particle is determined by only one parameter $f_0$; that is, the ratio of the Larmor radius at the flux rope axis to the flux rope radius. The analytical solutions show that cosmic rays cannot penetrate into the inner region of a flux rope by only gyration and gradient-curvature drift in the case of small $f_0$. Next, we perform a numerical simulation of a cosmic ray penetration into an interplanetary magnetic flux rope by adding small-scale magnetic field irregularities. The results show that cosmic rays can penetrate into a magnetic flux rope even in the case of small $f_0$ because of the effect of small-scale magnetic field irregularities. This simulation also shows that a cosmic ray density distribution is greatly different from that deduced from a guiding center approximation because of the effect of the finite Larmor radius and magnetic field irregularities for the case of a moderate to large Larmor radius compared to the flux rope radius.
\end{abstract}

\keywords{cosmic rays  --- diffusion --- interplanetary medium --- solar-terrestrial relations --- Sun: coronal mass ejections (CMEs) --- turbulence}

\notetoeditor{}

\section{Introduction}
\label{sec:intro}
Sudden decreases in cosmic ray intensity during geomagnetic storms were first observed by Forbush (1937) and are now well known as Forbush decreases. Observations of cosmic rays by neutron monitors and muon detectors revealed that Forbush decreases are caused by the passage of interplanetary shock waves with downstream highly turbulent magnetic fields and by magnetic clouds that have well-ordered strong magnetic fields. The observations also showed that there is an anisotropy in the cosmic ray distribution prior to and during a cosmic ray intensity decrease (Cane 2000 and references therein).\par

When interpreting and modeling the observations, it is very important to clearly separate the shock wave effect and magnetic cloud effect because their characteristics are quite different (Wibberenz \etal 1998). There are many theoretical and numerical models for the shock wave effect (e.g., Nishida 1982; Thomas \& Gall 1984; Chih \& Lee 1986; le Roux \& Potgieter 1991). Although data analyses of the magnetic cloud effect have also been carried out (e.g., Sanderson \etal 1990; Ananth \& Venkatesan 1993; Cane \etal 1994; Ifedili 2004; Subramanian \etal 2009), only a few theoretical and numerical models of the effect have been studied (Cane \etal 1995; Munakata \etal 2006; Kuwabara \etal 2009), and this has limited our understanding of the behavior of cosmic rays inside magnetic clouds.\par

Cane \etal (1995) (see also Wibberenz \etal 1998; Cane \etal 2000) modeled a cosmic ray penetration into a magnetic cloud by cross-field diffusion. They expressed the penetration by means of diffusion equation in a static vacuum cylinder region and deduced the cross-field diffusion coefficient for $\approx 1$ GV cosmic ray protons. Munakata \etal (2003) and Kuwabara \etal (2004) applied the muon observation data to deduce magnetic cloud geometry. These were the first derivations of the three-dimensional geometry of magnetic clouds from muon observation data. They assumed that the cosmic ray density distribution inside the magnetic cloud is Gaussian. Validity of the Gaussian distribution was not mentioned in the paper although the results were in good agreement with the geometry deduced from interplanetary magnetic field data. Munakata \etal (2006) and Kuwabara \etal (2009) obtained an improved, more realistic density distribution, derived from the Fokker--Planck equation for an isotropic phase-space density. This model used an expanding vacuum cylinder as a magnetic cloud model. From the analysis, they deduced the cross-field diffusion coefficient in an interplanetary magnetic cloud for high-rigidity cosmic rays observed by muon detectors.\par

Recently, the role of gradient-curvature drift for cosmic ray penetration into interplanetary magnetic clouds, which are often observed as magnetic flux ropes, was investigated by Krittinatham \& Ruffolo (2009). They developed an analytical model of the global magnetic field structure of a curved interplanetary magnetic flux rope with its roots connecting to the Sun. Cosmic ray particles guiding center orbits were traced in the model. It was shown that gradient-curvature drift plays a significant role for 1 GeV cosmic ray penetration into the magnetic flux rope. They emphasized that pitch angle scattering and perpendicular diffusion are also important.\par

All the models described above only considered guiding center motion and ignored the gyration of cosmic ray particles, namely, the Larmor radius of the cosmic ray particle is assumed to be infinitely small compared with the flux rope radius. However, in the case of high-rigidity cosmic rays inside an interplanetary magnetic flux rope, the assumption is not always valid. For example, the rigidity of cosmic rays detected by muon detectors is roughly 60 GV (e.g., Kuwabara \etal 2004). Assuming a typical magnetic flux rope near the Earth with a radius of 0.1 AU and an axial magnetic field intensity of 20 nT (Marubashi 1997), the ratio of the Larmor radius to the flux rope radius is then about 0.669, which is not much less than unity. This implies that the assumption described above is not valid for the high-rigidity cosmic rays inside an interplanetary magnetic flux rope. Therefore, we should examine the effect of a finite Larmor radius on cosmic ray penetration into an interplanetary magnetic flux rope.\par

The purpose of this article is to discuss the effect of a finite Larmor radius on the penetration process for cosmic rays into an interplanetary magnetic flux rope from the outside by considering the gyration of cosmic ray particles. In Section \ref{sec:analy}, we derive equations describing the behavior of cosmic ray particles inside a magnetic flux rope without magnetic field irregularities on the basis of the Newton-Lorentz equation of each cosmic ray particle. The numerical simulations of the cosmic ray penetration into the magnetic flux rope with magnetic field irregularities are described in Section \ref{sec:numer}. Sections \ref{sec:discussion} and \ref{sec:conclusions} are devoted to the discussion and the conclusions, respectively.\par

\section{The Model}
\subsection{Analytical Model}
\label{sec:analy}

According to pitch angle scattering theory, a charged particle is the most effectively scattered by a magnetic field irregularity having a spatial scale comparable to the Larmor radius of the particle. As discussed in Section \ref{sec:intro}, the Larmor radius of a high-rigidity cosmic ray particle, which can be observed using a muon detector, is comparable to the spatial scale of a magnetic flux rope near the Earth. This means that the high-rigidity cosmic rays inside the flux rope will not be scattered effectively by a small-scale magnetic field irregularity, which has a spatial scale much less than that of the flux rope. For this reason, we ignore cosmic ray scattering by such an irregularity in the analytic model. This model can exclude an effect of diffusion and consider only effects of gyration and gradient-curvature drift.\par

Generally, an interplanetary magnetic flux rope expands as it propagates through interplanetary space. However, we ignore flux rope expansion because its rate is many orders of magnitude less than the speed of a cosmic ray particle. We consider a non-expanding cylindrical magnetic flux rope model with radius $R_0$ (e.g., Marubashi 1997). The model is defined in cylindrical coordinates ($r$,$\varphi$,$z$) as
\begin{equation}
	\vec B = B_\varphi \vec e_\varphi + B_z \vec e_z,
	\label{B_vec}
\end{equation}
\begin{equation}
	B_\varphi=sB_0J_1(a\rho),
	\label{B_p}
\end{equation}
\begin{equation}
	B_z=B_0J_0(a\rho),
	\label{B_z}
\end{equation}
\begin{equation}
	\rho=\frac{r}{R_0}\quad (0\leq\rho\leq 1),
	\label{rho_def}
\end{equation}
where $B_0$ is the magnetic field intensity along the flux rope axis, and $J_0$ and $J_1$ are zeroth-order and first-order Bessel functions of the first kind, respectively; $s=1$ and $s=-1$ also correspond to parallel and antiparallel types of flux rope, respectively. Parallel/antiparallel flux ropes are those with electric current flowing parallel/antiparallel to the magnetic field in the flux rope. The constant $a\approx 2.40483$ is the smallest positive number of the zero point of the zeroth-order Bessel function of the first kind; that is, $J_0(a)=0$.\par

The Newton-Lorentz equation for a cosmic ray particle inside a flux rope is written as
\begin{equation}
	m\gamma\frac{d\vec v}{dt}=q\frac{\vec v}{c}\vec\times\vec B,
	\label{eom}
\end{equation}
where $m$, $q$, $\vec v$, $\gamma$, and $c$ are the mass, electric charge, velocity, Lorentz factor of a cosmic ray particle, and the speed of light, respectively. A solution of Equation (\ref{eom}) is generally described by one parameter $t$. However, as the solution mathematically defines a curved line on the spherical surface in the velocity space, it is possible to choose another parameter as long as one-to-one correspondence between the parameter and the time $t$ is confirmed. Since a distance $r$ from the flux rope axis to the particle position is described as

\begin{equation}
	r=\int_{t_0}^t v_r dt+r_0,
	\label{def_r}
\end{equation}
where $t_0$ and $r_0$ are constants, $r$ is a monotonous function of time $t$ and there is one-to-one correspondence between $r$ and time $t$ as long as the sign of $v_r$ does not change. If the sign of $v_r$ changes, we divide the curved line at the time of $v_r=0$ into separate segments. For each segment, there is one-to-one correspondence between $r$ and time $t$ because the sign of $v_r$ does not change. As a result, $r$ instead of the time $t$ can be applied as a parameter for each segment of the curved line. We can rewrite Equation (\ref{eom}) for each segment as an ordinary differential equation of $r$ or $\rho$ (Equation (\ref{rho_def})) by using the relation $dr=v_rdt$ derived from Equation (\ref{def_r}) as

\begin{equation}
	u_r\frac{d\vec u}{d\rho}=\frac{\vec u\vec\times\vec b}{f_0},
	\label{eom_r}
\end{equation}
where $\vec u$ and $\vec b$ are the normalized velocity and magnetic field, respectively, defined as

\begin{equation}
	\vec u=\vec v/\left|\vec v\right|,\quad \vec b=\vec B/B_0.
\end{equation}
$f_0$ is defined as
\begin{equation}
	f_0=\frac{m\gamma \left|\vec v\right|c}{qB_0R_0}=\frac{R_L}{R_0},
	\label{f_0}
\end{equation}
where $R_L$ is the Larmor radius of a cosmic ray particle for the magnetic field at the flux rope axis. Here, $f_0$ corresponds to the ratio of the Larmor radius of a cosmic ray particle to the flux rope radius. As the normalized magnetic field in the flux rope $\vec b$ depends only on $r$ or $\rho$, all segments of the curved line are described by the same equation (Equation (\ref{eom_r})). Also as two consecutive segments pass the same point in the phase space at the time of $v_r=0$, the solutions of Equation (\ref{eom_r}) for all segments should be the same in velocity space. Therefore, in the flux rope, the solution of Equation (\ref{eom_r}) together with a condition $\left|\vec u\right|=1$ and the solution of Equation (\ref{eom}) depict the same curved line in the velocity space. Equation (\ref{eom_r}) shows that the behavior of cosmic rays inside the flux rope is regulated only by parameter $f_0$.\par

\begin{figure}
	\plotone{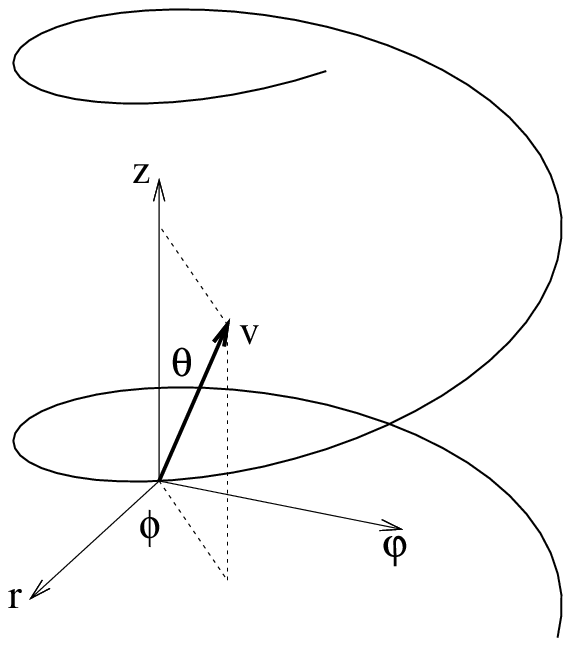}
	\caption{Definition of the coordinate system for the cosmic ray velocity $\vec v$ in the interplanetary magnetic flux rope. The origin is located on the cosmic ray particle. Here, $\theta$ is measured from the $z$ axis, which is parallel to the flux rope axis, while $\phi$ is measured from the $r$ axis, which leads from the center of the flux rope to the particle. The spiral line represents a magnetic field line of the flux rope.}
	\label{coord}
\end{figure}

The $r$, $\varphi$ and $z$ components of Equation (\ref{eom_r}) are expressed as
\begin{equation}
	u_r\frac{du_r}{d\rho}-\frac{u_\varphi^2}{\rho}=\frac{u_\varphi b_z-u_zb_\varphi}{f_0},
	\label{eqr}
\end{equation}
\begin{equation}
	\frac{du_\varphi}{d\rho}+\frac{u_\varphi}{\rho}=-\frac{b_z}{f_0},
	\label{eqp}
\end{equation}
and
\begin{equation}
	\frac{du_z}{d\rho}=\frac{b_\varphi}{f_0}.
	\label{eqz}
\end{equation}
The normalized velocity $\vec u$ can be written in spherical coordinates ($\theta$,$\phi$) as

\begin{equation}
	u_r=\sin\theta\cos\phi,\quad u_\varphi=\sin\theta\sin\phi,\quad u_z=\cos\theta,
	\label{def_u}
\end{equation}
where $\theta$ and $\phi$ are measured from the $z$ axis and $r$ axis, respectively (see Figure \ref{coord}). From the definition of the flux rope, the normalized magnetic field components $b_\varphi$ and $b_z$ are 

\begin{equation}
	b_\varphi=sJ_1(a\rho),\quad b_z=J_0(a\rho),
\end{equation}
respectively.\par

We set the boundary conditions as $u_r=\sin\theta_0\cos\phi_0$, $u_\varphi=\sin\theta_0\sin\phi_0$, and $u_z=\cos\theta_0$ at $\rho=\rho_0\le 1$. Equations (\ref{eqr}), (\ref{eqp}), and (\ref{eqz}) can be solved analytically. The solutions are expressed as
\begin{equation}
	u_r^2=1-\frac{1}{\rho^2}\left(\rho_0\sin\theta_0\sin\phi_0-C_\varphi\right)^2-\left(\cos\theta_0-C_z\right)^2,
	\label{sol_r}
\end{equation}
\begin{equation}
	u_\varphi=\frac{\rho_0\sin\theta_0\sin\phi_0-C_\varphi}{\rho},
	\label{sol_p}
\end{equation}
and
\begin{equation}
	u_z=\cos\theta_0-C_z,
	\label{sol_z}
\end{equation}
where
\begin{equation}
	C_\varphi=\frac{\rho J_1(a\rho)-\rho_0J_1(a\rho_0)}{af_0}
	\label{cp}
\end{equation}
and
\begin{equation}
	C_z=\frac{s\left[J_0(a\rho)-J_0(a\rho_0)\right]}{af_0}.
	\label{cz}
\end{equation}
These solutions include gyration and gradient-curvature drift effects. Equation (\ref{sol_r}) shows that the particle position $\rho$ should satisfy the relation:
\begin{equation}
	(\rho_0\sin\theta_0\sin\phi_0-C_\varphi)^2+\rho^2(\cos\theta_0-C_z)^2-\rho^2\leq 0.
	\label{bdry}
\end{equation}
This means that only particles having the boundary condition $\theta_0$ and $\phi_0$ at $\rho_0$ can reach the region where Equation (\ref{bdry}) is satisfied.

\begin{figure}
	\plotone{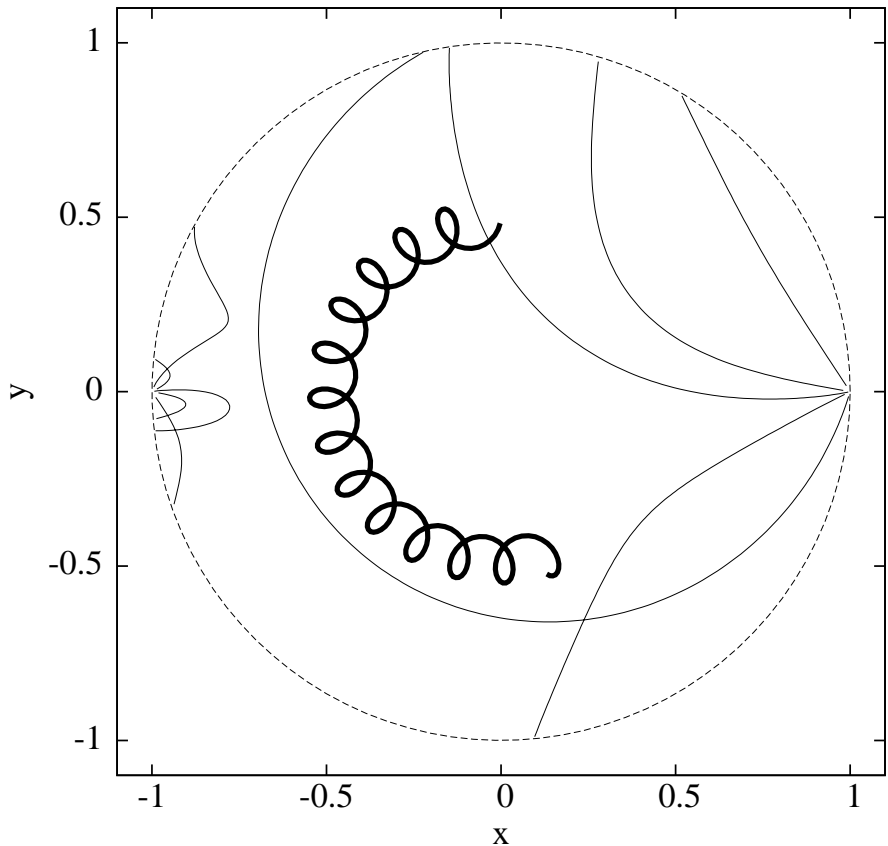}
	\caption{Particle trajectories on the $r-\varphi$ plane observed in the numerical simulation. The dashed circle with a radius of 1 is the edge of the flux rope. The solid curves starting from the point $(x,y)=(1,0)$ and $(x,y)=(-1,0)$ are the particle trajectories of 60 GV cosmic rays and 6 GV cosmic rays for various boundary conditions $\theta_0$ and $\phi_0$ at $\rho_0=1$, respectively. The thick spiral curve shows the trajectory of a 6 GV trapped cosmic ray particle (see text) for randomly selected boundary condition $\theta_0$ and $\phi_0$ at $\rho_0=0.5$. A 0.1 AU radius flux rope with a 20 nT axial magnetic field is used for typical values near the Earth.}
	\label{xy}
\end{figure}

To examine the validity of the transformation from Equation (\ref{eom}) to Equation (\ref{eom_r}), we compare the analytical solutions with a numerical simulation, in which a cosmic ray particle is directly traced in the magnetic flux rope. We integrate Equation (\ref{eom}) and $\vec v=d\vec r/dt$ by using the Buneman-Boris Method. Needless to say, the numerical simulation includes both gyration and gradient-curvature drift effects. Hereafter, we consider a parallel-type magnetic flux rope model (i.e., $s=1$) without loss of generality. We use $R_0=0.1$ AU and $B_0=20$ nT as typical values near the Earth (Marubashi 1997). Figure \ref{xy} shows the particle trajectories on the $r-\varphi$ plane. A dashed circle with a radius of 1 in the figure is the edge of flux rope. The solid curves starting from the point $(x,y)=(1,0)$ and $(x,y)=(-1,0)$ are the particle trajectories of 60 GV cosmic rays, which correspond to $f_0=0.669$, and those of 6 GV cosmic rays, which correspond to $f_0=0.0669$, respectively, for various boundary conditions $\theta_0$ and $\phi_0$ at $\rho_0=1$. The starting points are not important for the results because the magnetic field in the flux rope is symmetrical with regard to the $z$ axis. All these particles entering from the outside of the flux rope are escaping from the flux rope within one gyro motion. A thick spiral curve in the figure shows a trajectory of a 6 GV trapped cosmic ray particle for randomly selected boundary condition $\theta_0$ and $\phi_0$ at $\rho_0=0.5$. The trapped particle trajectory obviously shows a gradient-curvature drift. The drift velocity components are in the $\varphi$ and $z$ directions as calculated from the definition of the flux rope.\par

\begin{figure}
	\plotone{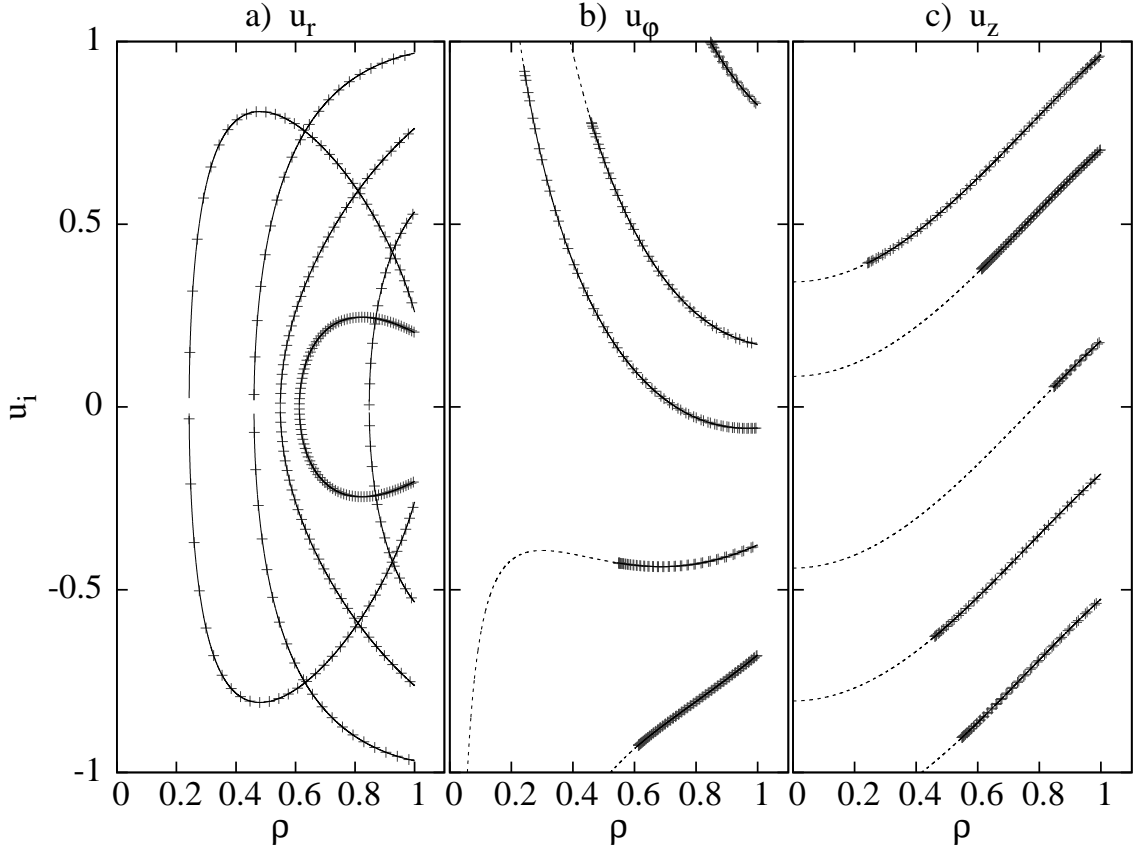}
	\caption{Three components of the normalized velocity for 60 GV cosmic ray particles for various penetration angles $\theta_0$ and $\phi_0$ at $\rho_0=1$, which correspond to those in Figure \ref{xy}. The components are depicted against the normalized distance from the flux rope axis. The edge of the flux rope is at $\rho=1$. Panels a), b), and c) show the $r$, $\varphi$, and $z$ components of the velocity, respectively. The solid lines are expressed as Equations (\ref{sol_r}), (\ref{sol_p}), and (\ref{sol_z}). Unphysical solutions, in which $u_r$ is an imaginary number, are also drawn by dashed lines. The gray plus signs show results of the numerical simulation.}
	\label{rvrvpvz_60GV}
\end{figure}

Figure \ref{rvrvpvz_60GV} shows the three components of normalized velocity $\vec u$ against the normalized distance $\rho$ from the flux rope axis for 60 GV cosmic rays. Panels a), b), and c) show the $r$, $\varphi$, and $z$ components, respectively. Gray plus signs show results of numerical simulation for various penetration angles $\theta_0$ and $\phi_0$ at $\rho_0=1$, corresponding to those in Figure \ref{xy}. Solid lines show the analytical solutions for the same parameters used in the numerical simulation. It is evident that the analytical solutions exactly trace the simulation results. It is noticed that unphysical solutions, in which $u_r$ is an imaginary number, are also drawn by dashed lines in Panels b) and c) in the figure.\par

\begin{figure}
	\plotone{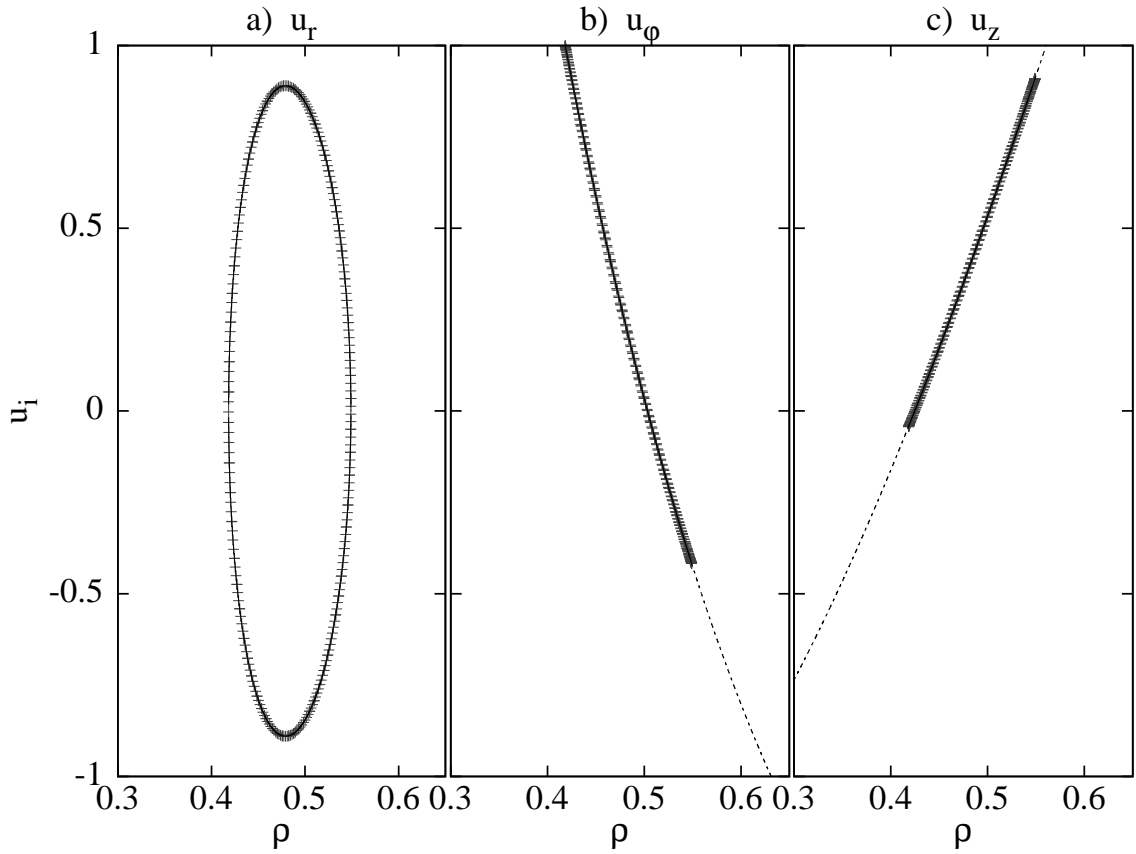}
	\caption{Three components of the normalized velocity for a 6 GV trapped cosmic ray particle for angle $\theta_0$ and $\phi_0$ at $\rho_0=0.5$, which correspond to those in Figure \ref{xy}. The components are depicted against the normalized distance from the flux rope axis. The solid and dashed lines, and gray plus signs, are the same as those in Figure \ref{rvrvpvz_60GV}.}
	\label{rvrvpvz_trap}
\end{figure}

Figure \ref{rvrvpvz_trap} also shows the three components of normalized velocity against the normalized distance for a 6 GV trapped cosmic ray particle. Gray plus signs show results of numerical simulation, represented by the spiral curve in Figure \ref{xy}. Solid lines are the analytical solutions for the same parameters used in numerical simulation. It is also evident that the analytical solutions exactly trace the simulation results. It is again noticed that unphysical solutions, in which $u_r$ is an imaginary number, are also drawn in Panels b) and c) in the figure. The trapped cosmic ray particle obviously gyrates and drifts in the magnetic fields, as described above. The trajectories of these particles in velocity space show oscillatory motions on the lines described by our analytical solutions. Figures \ref{xy}, \ref{rvrvpvz_60GV}, and \ref{rvrvpvz_trap} indicate that our analytical solutions represent exact particle motion including particle gyration and gradient-curvature drift in the magnetic flux rope.\par

Hereafter, we consider cosmic rays penetrating a flux rope from the outside; that is, $\theta_0$ and $\phi_0$ at $\rho_0=1$ as the boundary condition.
Not all particles reach the axis $\rho=0$ in Figure \ref{rvrvpvz_60GV} because only particles with specific angles $\theta_0$ and $\phi_0$ satisfy Equation (\ref{bdry}) at $\rho=0$. A cosmic ray particle entering the flux rope at angles $\theta_0$ and $\phi_0$ moves along the lines shown in the Panels b) and c) in Figure \ref{rvrvpvz_60GV} until it reaches the innermost point where the left-hand side of Equation (\ref{bdry}) is zero. The particle then returns along the same line to the edge of the flux rope ($\rho=1$). The $\varphi$ and $z$ components of the velocity of the escaping particle are the same as those of the entering particle if $\rho$ is the same, and the $r$ component of velocity is in the opposite direction and has the same magnitude when the particle is entering and escaping. Panel c) in Figure \ref{rvrvpvz_60GV} and Equation (\ref{sol_z}) show that the $z$ component of the velocity decreases when the particle is entering and increases when it is escaping. This is because the $z$ component of the Lorentz force is always negative (positive) when cosmic ray particles move from (to) the edge of the flux rope to (from) the innermost point. We find from Equation (\ref{cz}) that the slopes of the trajectories in the panel c) in Figure \ref{rvrvpvz_60GV} are steeper for lower-rigidity cosmic rays and less steep for higher-rigidity cosmic rays (not shown in the figure).\par

\begin{figure}
	\plotone{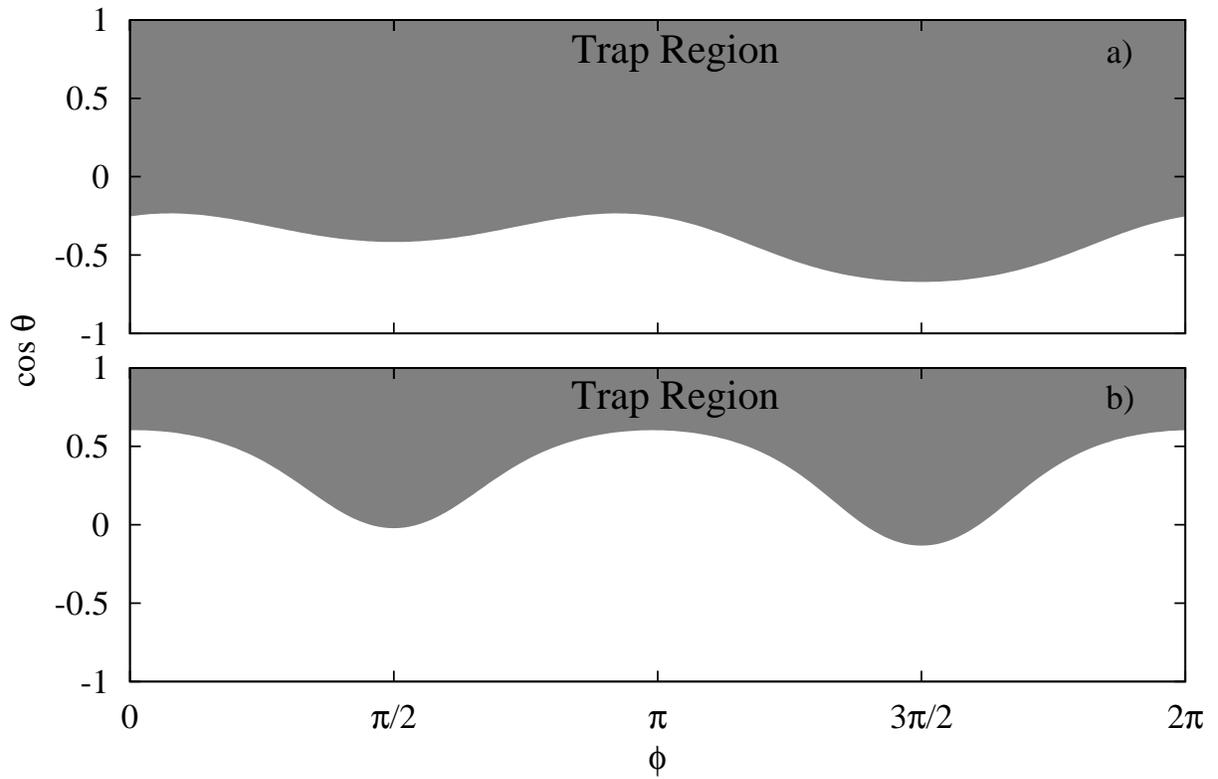}
	\caption{Trap region (gray) for $f_0=0.0669$. Panels a) and b) are the trap region for $\rho=0.85$ and $0.95$, respectively. The solid angle of the trap region depends on $\rho$.}
	\label{trap}
\end{figure}

Cosmic ray particles penetrating from the outside of a flux rope must have $u_r^2\geq 0$ at the edge of a flux rope, $\rho_0=1$. From this fact and Equations (\ref{def_u}) to (\ref{cz}), we find that the cosmic ray particles penetrating from the outside of the flux rope cannot exist in the region that satisfies 
\begin{equation}
	1-(\rho\sin\theta\sin\phi+C_\varphi)^2-(\cos\theta+C_z)^2 < 0,
	\label{Dregion}
\end{equation}
when we call it a trap region. Figure \ref{trap} shows the trap region for $f_0=0.0669$. Panels a) and b) are the trap region for $\rho=0.85$ and $0.95$, respectively. These panels show that the solid angle of the trap region depends on $\rho$. The solid and dashed lines in Figure \ref{trap_area} show $\rho$ dependence of the solid angle of the trap region normalized by $4\pi$ steradian for $f_0=0.0669$ and $f_0=0.669$, respectively. The solid angle of the trap region is larger (smaller) for smaller (larger) $\rho$ for a given $f_0$, and is $4\pi$ steradian in the inner region of the flux rope for $f_0=0.0669$. If cosmic ray particles are in the trap region, the particles cannot escape from the flux rope and are trapped.\par

We calculate the forbidden region to which a cosmic ray particle cannot arrive from the outside of the flux rope for any $\theta_0$ and $\phi_0$ using Equation (\ref{bdry}).
Figure \ref{penet} depicts the forbidden region. The abscissa and ordinate are $f_0$ and $\rho$, respectively. The figure shows that low-rigidity cosmic rays, which can be observed by a neutron monitor, usually cannot arrive at the inner region of the flux rope. For example, in the case of $f_0=0.0669$, which corresponds to a 6 GV cosmic ray particle, the particle cannot penetrate the region for $\rho\leq 0.769$. Thus, the low-rigidity cosmic rays cannot reach the inner region of a magnetic flux rope by only gyration and gradient-curvature drift.\par

However, many events of cosmic ray decreases observed by neutron monitors suggest that cosmic rays exist in the inner regions of flux ropes. Therefore, other mechanisms for penetration into the inner region of flux rope by low-rigidity cosmic ray particles are required. For these particles, $f_0$ is much less than unity and they can be scattered effectively by small-scale magnetic field irregularities in the flux rope. For this reason, we consider small-scale magnetic field irregularities in the interplanetary magnetic flux ropes in Section \ref{sec:numer}.\par

On the other hand, since the Larmor radius of high-rigidity cosmic ray particles is comparable to the flux rope radius (e.g., $f_0=0.669$ for 60 GV), these particles can reach the inner region of the flux rope from the outside by gyration. While the high-rigidity cosmic rays can reach the inner region of the magnetic flux rope by only gyration, the effect of the small-scale magnetic field irregularities on the cosmic ray penetration into the flux rope is not clear. Therefore, we also investigate the effect of the small-scale magnetic field irregularities in the interplanetary magnetic flux ropes for high-rigidity cosmic rays in Section \ref{sec:numer}.\par

In this study, we only considered the case of a flux rope located around the Earth. Since a magnetic flux rope is ejected from the Sun and propagates through interplanetary space toward the Earth while expanding, we must consider the flux rope near the Sun. According to the expanding cylindrical force-free magnetic flux rope model (Shimazu \& Vandas 2002; Berdichevsky \etal 2003), the flux rope radius $R_0$ is directly proportional to the time $t$ that has passed since the flux rope ejection at the Sun, and the magnetic field intensity on the flux rope axis $B_0$ is inversely proportional to time $t$ squared. From these dependences of $R_0$ and $B_0$, and Equation (\ref{f_0}), the parameter $f_0$ is directly proportional to time $t$. This means that the parameter $f_0$ near the Sun is much less than that around the Earth; for example, $f_0=0.0669$ for a typical flux rope located 0.1 AU from the Sun and 60 GV cosmic ray particles. This is the same situation as having low-rigidity cosmic ray particles around the Earth because the behavior of cosmic rays inside a flux rope is regulated by only the parameter $f_0$. Therefore near the Sun, even high-rigidity cosmic rays cannot penetrate the interior of the flux rope by gyration alone, and the role of magnetic field irregularities must also be considered.\par

\subsection{Numerical Model}
\label{sec:numer}
To examine the role of magnetic field irregularities in the penetration of cosmic rays into interplanetary magnetic flux ropes, we perform numerical simulations in which cosmic ray particles are directly traced in the turbulent magnetic field. The direct simulations solving the Newton-Lorentz equation in the turbulent magnetic field were recently performed by some authors, for instance, in studies of cross-field diffusion (e.g., Micha\l ek \& Ostrowski 1997; Giacalone \& Jokipii 1999; Mace \etal 2000; Qin \etal 2002; Zimbardo \etal 2006), pitch angle diffusion (Qin \& Shalchi 2009), suppression of particle drift (Minnie \etal 2007), and solar energetic particle transport (Qin \etal 2004).\par

We trace the cosmic ray particles in the magnetic flux rope with the addition of magnetic field irregularities. Magnetic field irregularities are defined as the superposition of many sinusoidal waves,
\begin{equation}
	\vec B_{total}=\vec B+\delta\vec B,
	\label{BdB}
\end{equation}
\begin{equation}
	\delta\vec B=\sum_{n=1}^N \vec\zeta_nBC(k_n)\exp\left[i\left(\vec k_n\cdot\vec r+\psi_n\right)\right],
	\label{def_dB}
\end{equation}
\begin{equation}
	C(k_n)^2=\left[\frac{\delta B}{B}\right]^2F(k_n)\Delta k_n\left[\sum_{n=1}^N F(k_n)\Delta k_n\right]^{-1},
	\label{def_C}
\end{equation}
\begin{equation}
	F(k_n)=\left[1+\left(l_ck_n\right)^{5/3}\right]^{-1},
	\label{spectra}
\end{equation}
where $B=\left|\vec B\right|$, $\delta B=\left|\delta\vec B\right|$, and $k_n=\left|\vec k_n\right|$. Vector $\vec\zeta_n$ is perpendicular to $\vec k_n$ to assure $\mbox{div}\vec B_{total}=0$. Equation (\ref{spectra}) defines a Kolmogorov-type wave number spectrum of magnetic field irregularities. These definitions are similar to those of previous studies (e.g., Giacalone \& Jokipii 1999). The minimum and maximum wave number used in the simulation correspond to wave length of $2 R_0$ (diameter of flux rope) and $2\times10^{-4} R_0$, respectively. $l_c$ is a correlation length of the turbulence spectrum and is set as $0.5 R_0$. One sinusoidal wave is determined by four random numbers; two for a direction of wave number vector, one for a phase of wave, and the other for a polarization. The 255 sinusoidal waves are superimposed on the background magnetic field. We confirmed that these waves are sufficient to reproduce a diffusion process and that results are insensitive to the number of waves. We use a value $[\delta B/B]^2=0.01$ as an intensity of magnetic field irregularities in this study, which is consistent with values of a small percent found in some magnetic clouds (e.g., Subramanian \etal 2009).\par

In the simulation, the background magnetic field $\vec B$ is the flux rope magnetic field defined as Equations (\ref{B_vec}) to (\ref{rho_def}). We consider a parallel-type magnetic flux rope model with the typical values of $R_0=0.1$ AU and $B_0=20$ nT near the Earth. For a time dependence of flux rope radius and magnetic field, we use the model by Shimazu \& Vandas (2002), which can very effectively reproduce observations. In this model, when the flux rope is located near the Sun, for instance, at 0.1 AU from the Sun, $R_0$ and $B_0$ will be 0.01 AU and 2,000 nT, respectively.\par

Cosmic ray particles are traced by the Buneman-Boris Method. The number of particles $M$ is 10,000. All these particles are initially located at the flux rope edge with randomly distributed penetration angles. When a particle escapes from the edge, a new particle is set up immediately at the edge and we trace its orbit. A different set of the random numbers for magnetic field irregularities are selected for each particle.\par

\begin{figure}
	\plotone{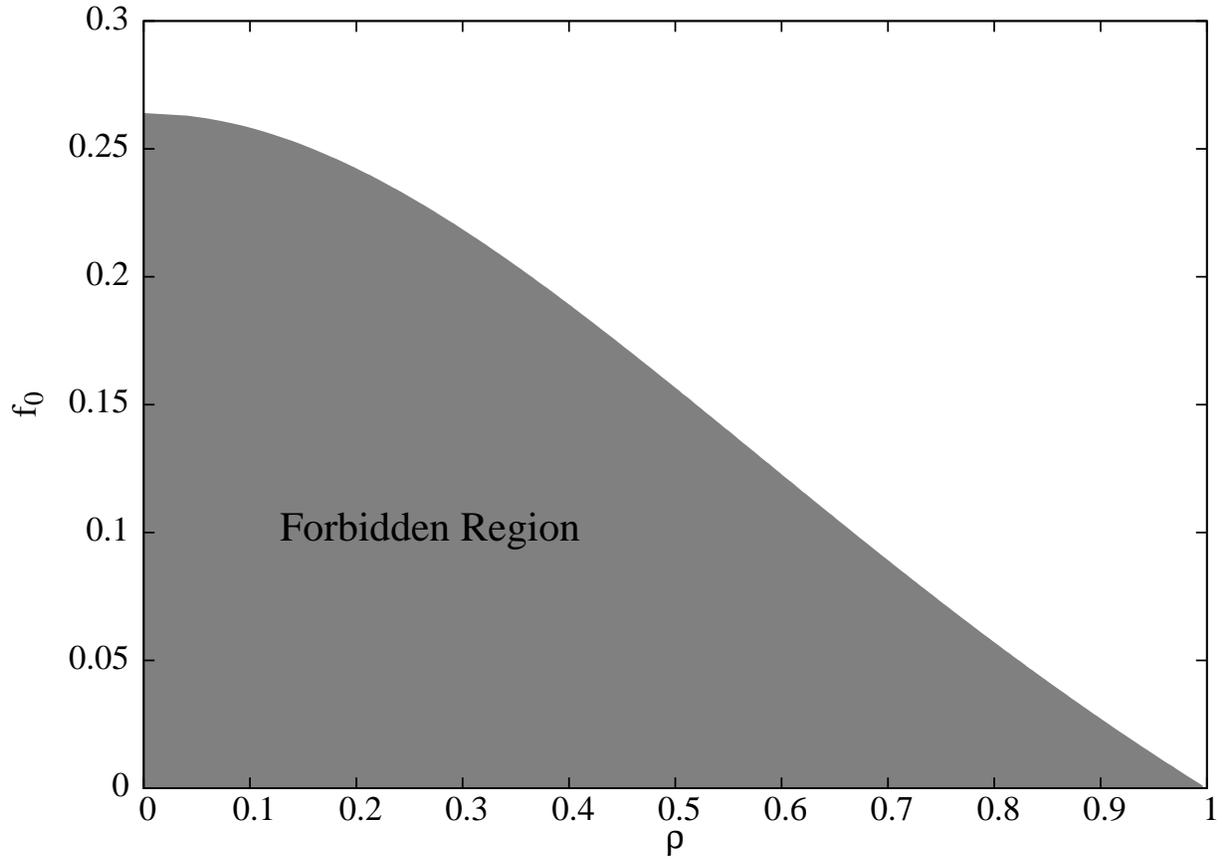}
	\caption{Forbidden region (gray) where cosmic rays penetrating from the outside cannot enter. The abscissa and ordinate are, respectively, the parameter $f_0$ and distance from the flux rope axis normalized by the flux rope radius.}
	\label{penet}
\end{figure}

\begin{figure}
	\plotone{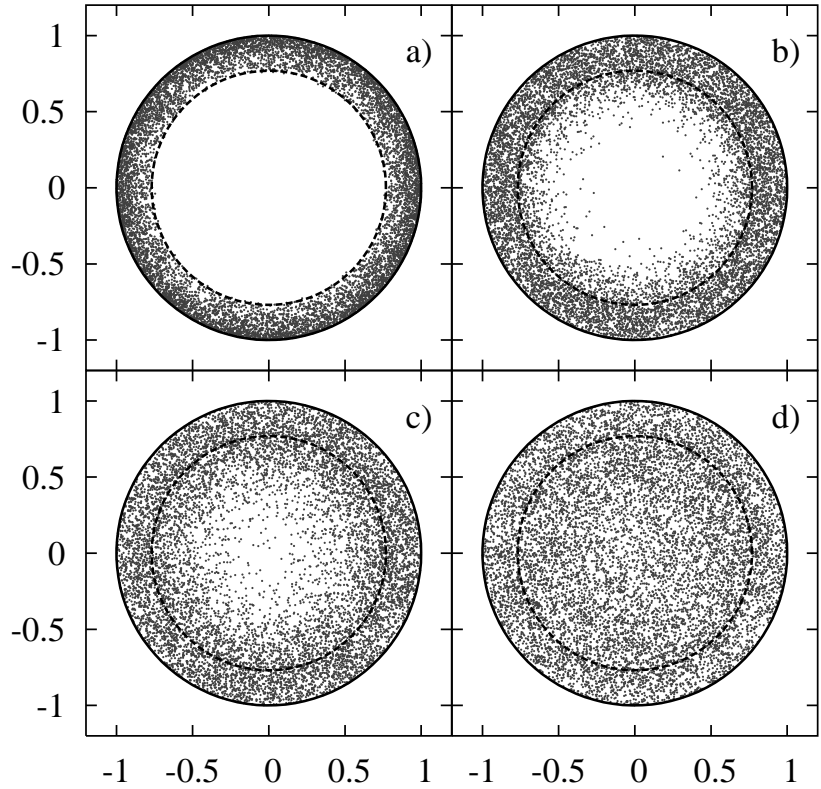}
	\caption{Cosmic ray particle distribution on the $r-\varphi$ plane for the case that the flux rope is located at 0.1 AU from the Sun and that the cosmic ray rigidity is 60 GV. Solid and dashed circles show the flux rope edge and the boundary of the forbidden region, respectively. Panels a), b), c), and d) are for the time $t=$ 2.5, 250, 750, and 2,500 ($\omega_p^{-1}$).}
	\label{cospos_60GV010}
\end{figure}

Figure \ref{cospos_60GV010} depicts the simulated particles on the $r-\varphi$ plane for a case where the flux rope is located at 0.1 AU from the Sun and the cosmic ray rigidity is 60 GV. In this case, the parameter $f_0$ is 0.0669. Solid and dashed circles in the figure show the flux rope edge and the boundary of the forbidden region shown in Figure \ref{penet}, respectively. The gray tiny points are cosmic ray particles. Panels a), b), c), and d) are snapshots for the time $t=$ 2.5, 250, 750, and 2,500 $\omega_p^{-1}$, where $\omega_p$ is the proton cyclotron frequency ($\omega_p=0.479$ sec$^{-1}$ in this study). Panel a) shows that almost all cosmic ray particles are located outside of the forbidden region and the effect of magnetic field irregularities is small. However, after approximately nine minutes later, in Panel b), the effect of magnetic field irregularities clearly appears, and a large number of particles penetrate into the forbidden region in the flux rope. In Panel c), some particles reach the center of the flux rope. In Panel d), cosmic ray particle distribution is nearly uniform. Penetration into the forbidden region is due not to drift motion but to the diffusion process caused by particle scattering and magnetic field line random walk (Jokipii \& Parker 1968).\par

\begin{figure}
	\plotone{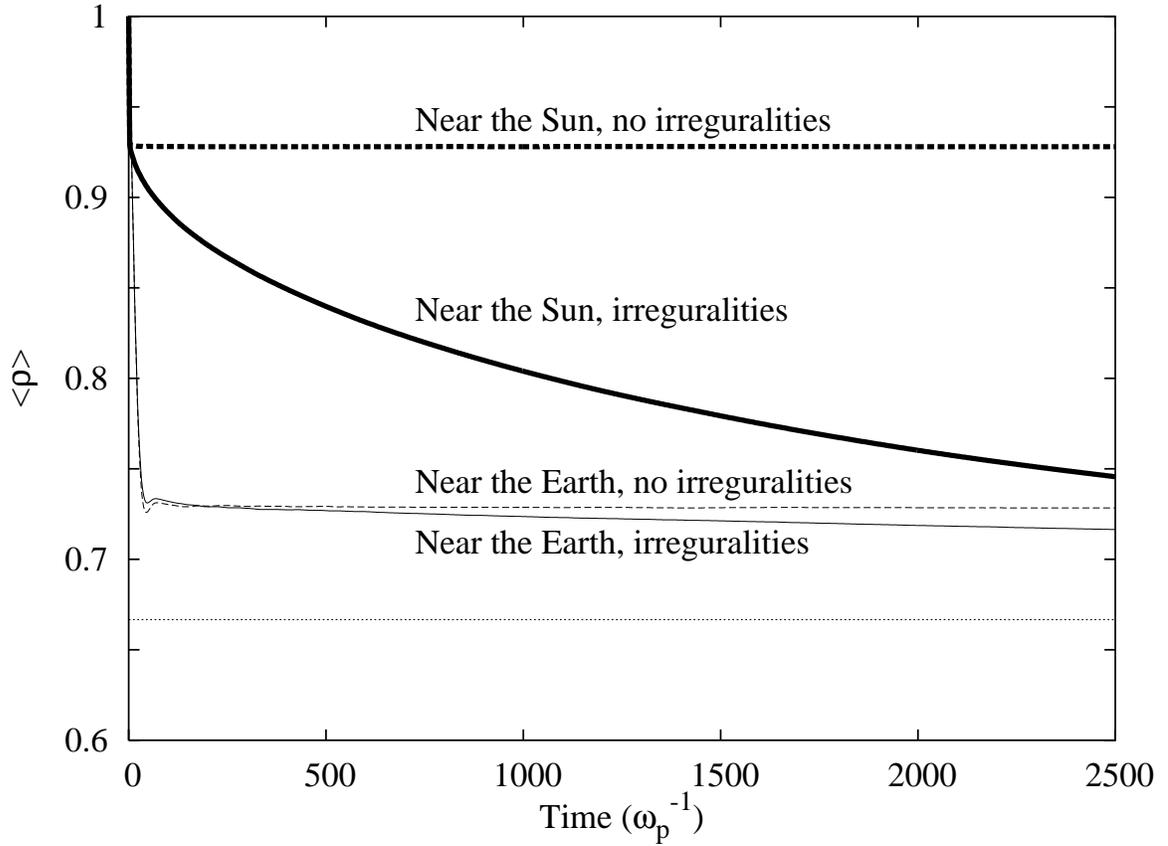}
	\caption{A mean radial distance calculated from the simulation results. Thick and thin lines are the results for the case near the Sun and Earth, respectively. Solid and dashed lines are the results with and without magnetic field irregularities, respectively. The dotted horizontal line is a mean radial distance for the uniformly distributed case.}
	\label{rm_sim}
\end{figure}

Figure \ref{rm_sim} shows a time evolution of a mean radial distance of 60 GV cosmic ray particles in the typical flux rope, whose parameters are $R_0=0.1$ AU and $B_0=20$ nT near the Earth. The mean radial distance at time $t$ is calculated as
\begin{equation}
	\left<\rho\right>=\frac{1}{M(k+1)}\sum_{j=0}^{k}\sum_{i=1}^{M}\rho_{ij},
	\label{rho_k}
\end{equation}
where $k$ is simulation time step number until time $t$ and $\rho_{ij}$ is a distance from the flux rope axis of the $i$th cosmic ray particle at the simulation time step $j$. In this definition, the entire simulation time step up to time $t$ is used to calculate $\left<\rho\right>$. This definition is based on a continuous cosmic ray penetration from the outside of the flux rope. The thin dotted horizontal line in Figure \ref{rm_sim} shows the result if the particles are uniformly distributed in the flux rope, $\left<\rho\right>=2/3$. Thick and thin dashed lines show the cases without magnetic field irregularities near the Sun ($f_0=$ 0.0669) and Earth ($f_0=$ 0.669), respectively. Thick and thin solid lines show the cases with magnetic field irregularities for near the Sun and Earth, respectively. We find from the figure that the mean radial distance immediately reaches steady state values in all cases. In the case near the Sun (thick solid line), the mean radial distance does obviously decrease and is asymptotic to the uniform distribution line as the time elapses. This means that the cosmic ray particles penetrate into the inner region of the flux rope by diffusion, as shown in Figure \ref{cospos_60GV010}. A mean radial distance for the case near the Earth (thin lines) is already small at the beginning of the calculation compared to that for the case near the Sun. This is because cosmic ray particles are able to reach the inner region of flux rope by gyration because of the large $f_0$ parameter. However, the decrease of the mean radial distance is small even if the time elapses. We find from the figure that it takes more time to penetrate into the flux rope by diffusion; thus the gyration is more effective than the diffusion for high $f_0$ such as near the Earth. The diffusion is effective when the flux rope is located near the Sun (small $f_0$).\par

\begin{figure}
	\plotone{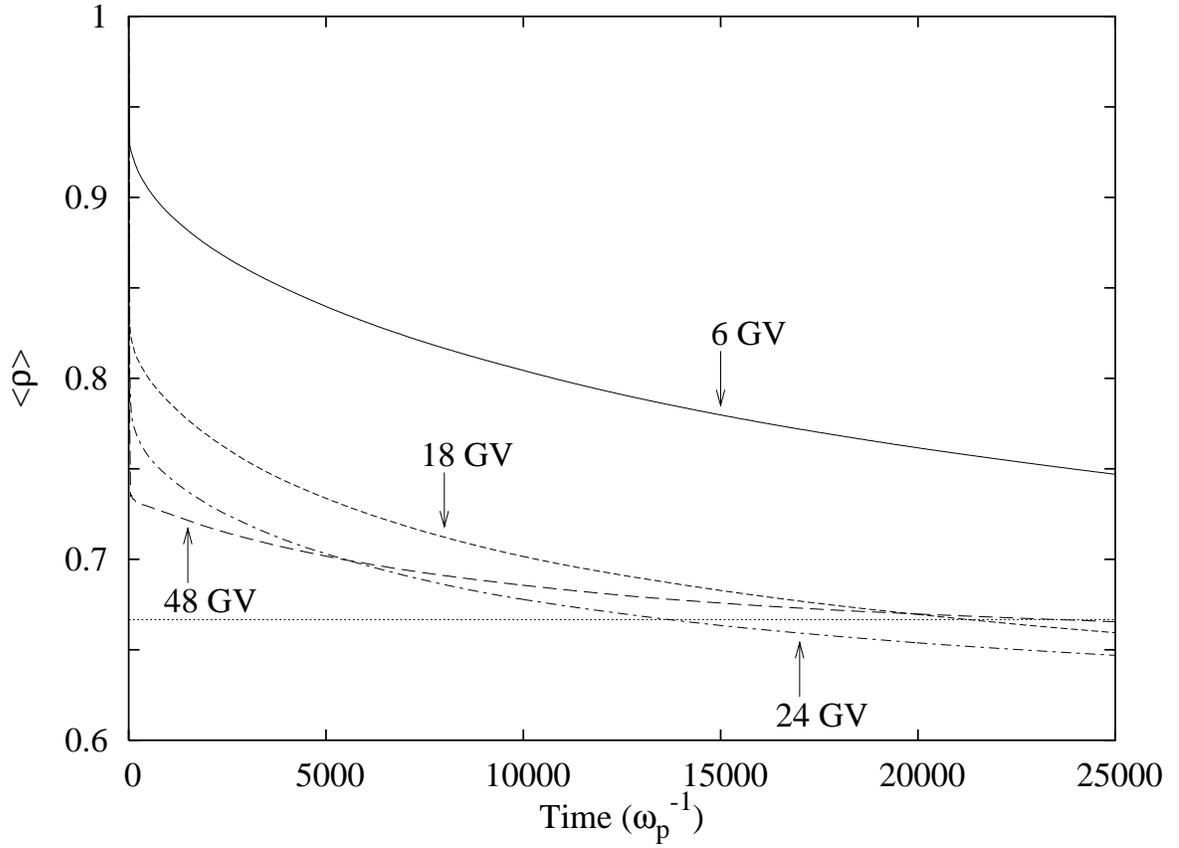}
	\caption{A mean radial distance calculated from the simulation results for the case of near the Earth. Solid, dashed, dash-dotted, and long-dashed lines are the results for 6, 18, 24, and 48 GV cosmic rays, respectively. Flux rope parameters are $R_0=0.1$ AU and $B_0=$ 20 nT. The dotted horizontal line is the same as in Figure \ref{rm_sim}.}
	\label{rm25000}
\end{figure}

Figure \ref{rm25000} shows the mean radial distance for 6, 18, 24, and 48 GV cosmic ray particles for the flux rope located near the Earth. The simulation run time is 10 times longer than that in the previous case shown in Figure \ref{rm_sim}. The solid, dashed, dash-dotted, and long-dashed lines are the results for 6, 18, 24, and 48 GV cosmic ray particles, which correspond to $f_0=$ 0.0669, 0.201, 0.268, and 0.535, respectively. The dotted horizontal line shows the mean radial distance for which the particles are uniformly distributed in the flux rope. The mean radial distances for all cases clearly decrease with time. We find from comparison between the thick solid line in Figure \ref{rm_sim} and the solid line in Figure \ref{rm25000} that cosmic ray particles attain the same distribution in the flux rope if the parameter $f_0$ is the same even when the magnetic field irregularities exist. The time taken to attain this distribution is proportional to the flux rope radius. This fact means that some distribution of the cosmic ray density can be attained in a short time when the flux rope is located near the Sun. It is interesting that the mean radial distances drop under the uniform distribution line in all cases except for the 6 GV case. This means that a density in the inner region of a flux rope becomes larger than the density in outer region. This result shows the significance of the effect of finite Larmor radius because this kind of density distribution cannot be attained by only the diffusion process.\par

\begin{figure}
	\plotone{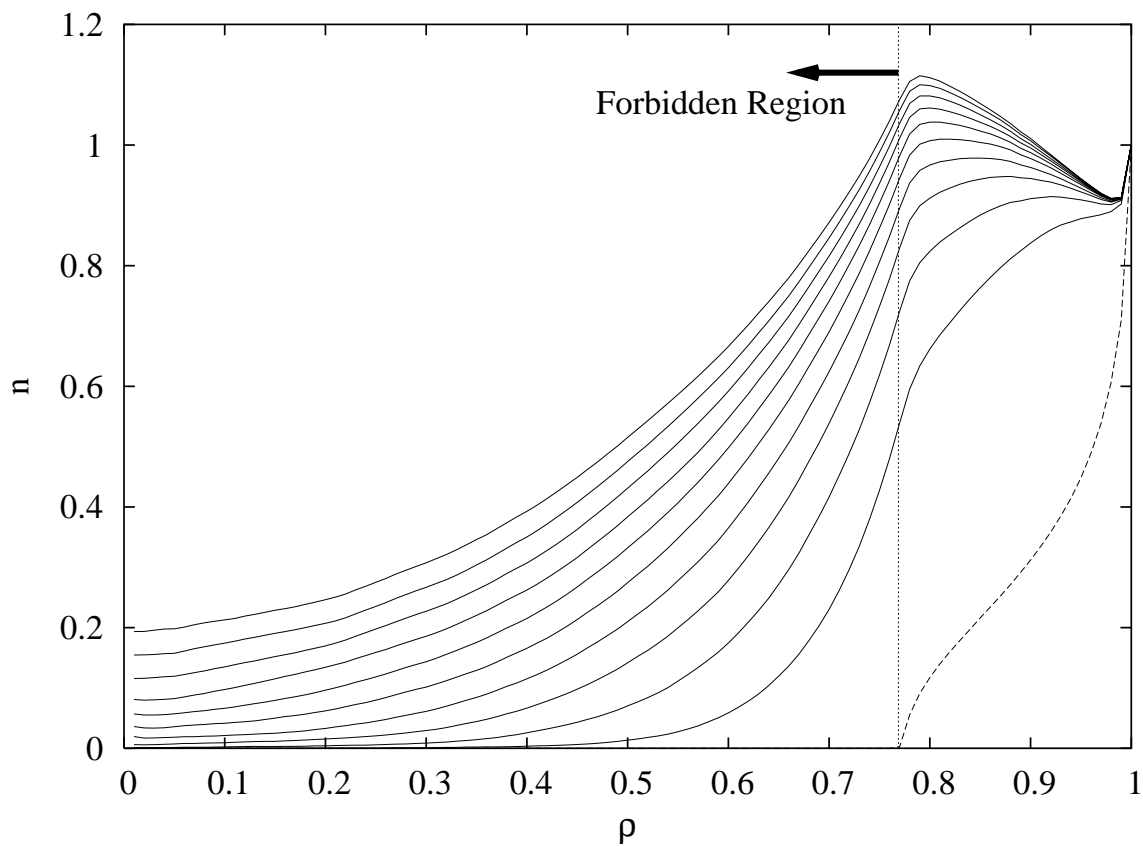}
	\caption{Cosmic ray density distributions for 6 GV cosmic rays calculated by simulations. Flux rope parameters are the same as the ones used in Figure \ref{rm25000}. The dashed line is the distribution for the case without magnetic field irregularities. Ten solid lines from the bottom to the top show the time evolution of the density distribution for the case with magnetic field irregularities. The region inside the dotted vertical line shows the forbidden region calculated from the analytical model. The density $n$ is normalized to unity at the edge of the flux rope.}
	\label{dhist6}
\end{figure}

\begin{figure}
	\plotone{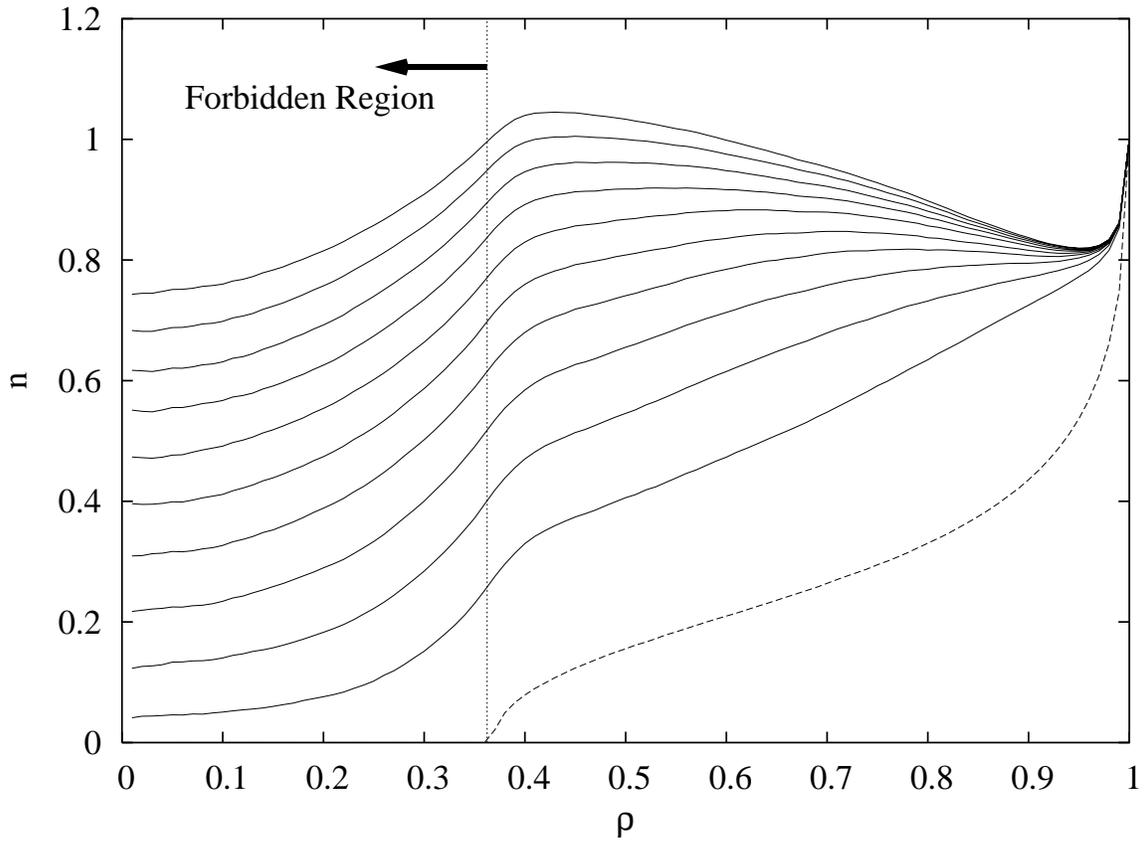}
	\caption{Same as Figure \ref{dhist6}, except the cosmic ray rigidity is 18 GV.}
	\label{dhist18}
\end{figure}

\begin{figure}
	\plotone{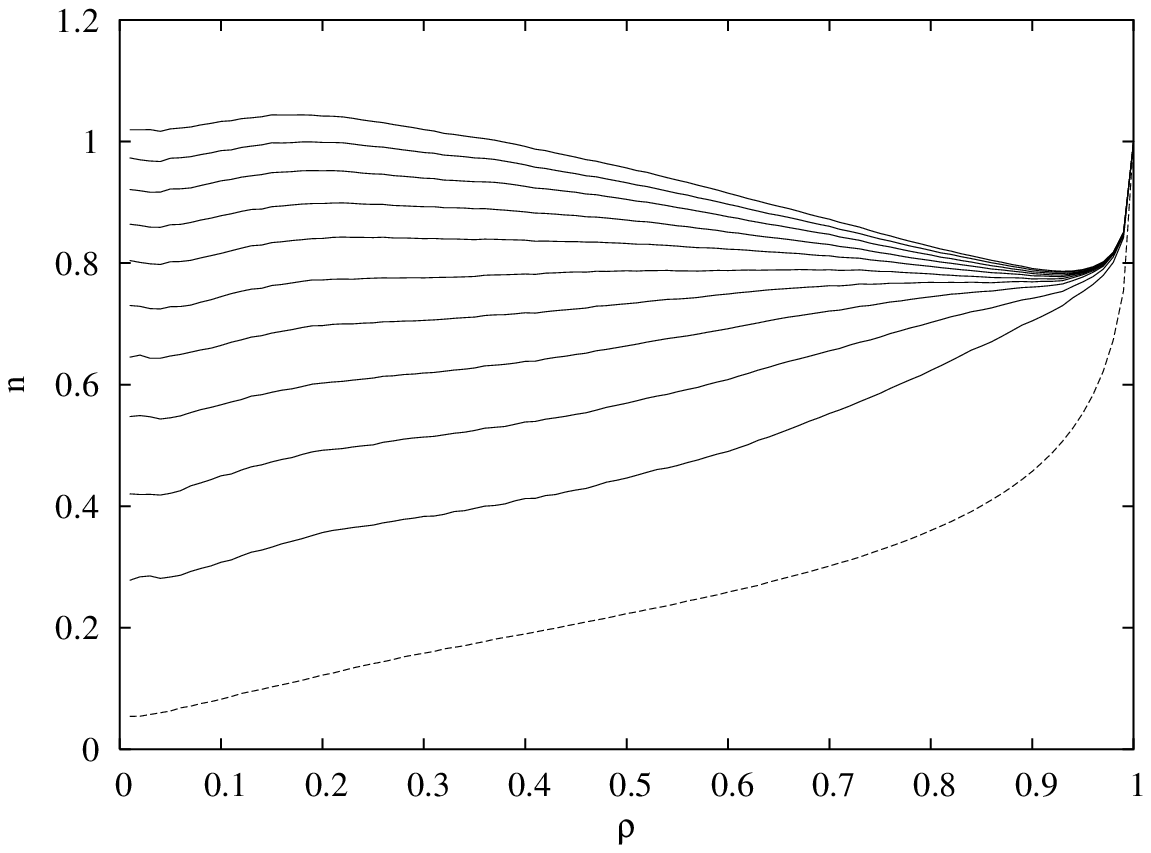}
	\caption{Same as Figure \ref{dhist6}, except the cosmic ray rigidity is 24 GV. There is no forbidden region for this rigidity cosmic ray particle.}
	\label{dhist24}
\end{figure}

\begin{figure}
	\plotone{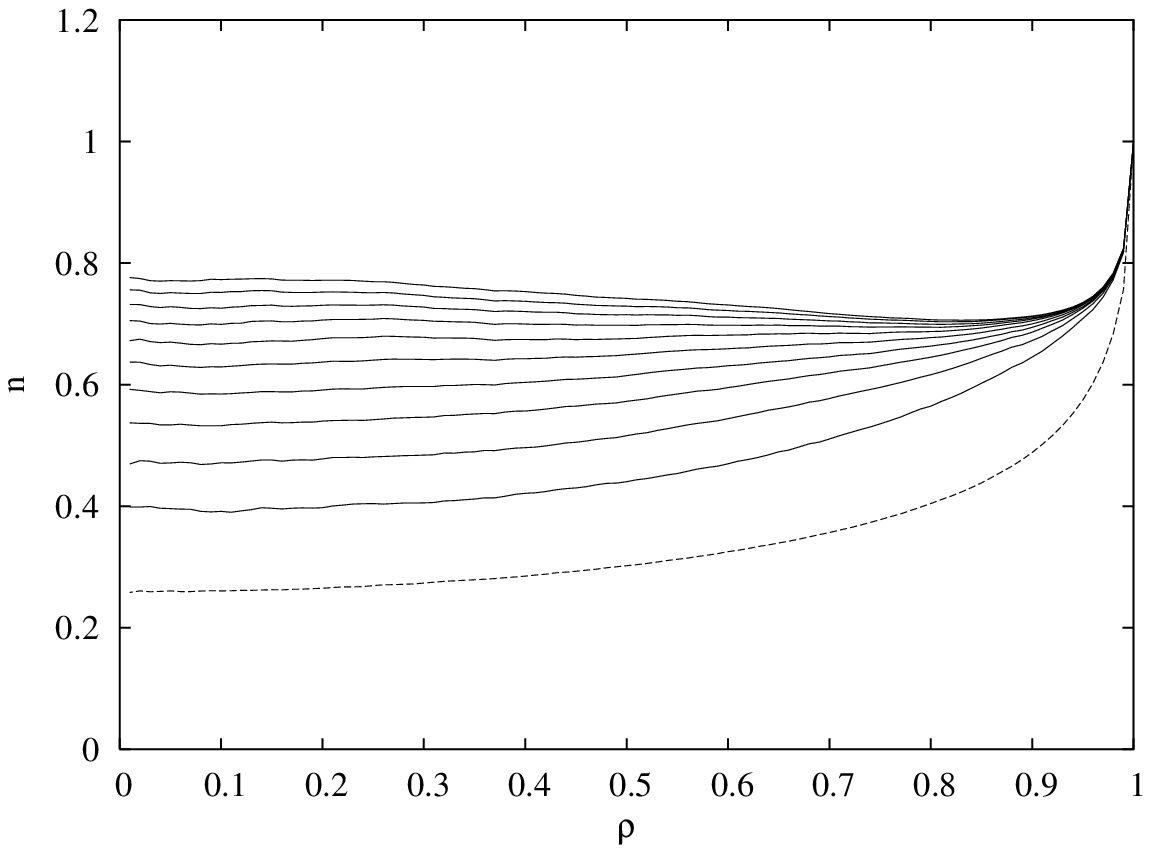}
	\caption{Same as Figure \ref{dhist6}, except the cosmic ray rigidity is 48 GV. There is no forbidden region for this rigidity cosmic ray particle.}
	\label{dhist48}
\end{figure}

Figures \ref{dhist6} to \ref{dhist48} show the calculated cosmic ray radial density distributions inside the flux rope located near the Earth for the cases of 6, 18, 24, and 48 GV cosmic ray particles, respectively. All the simulation time steps up to time $t$ are used to calculate the density distributions at time $t$. The dashed lines in the figures are density distributions in the steady state, which are calculated by simulations without magnetic field irregularities. Ten solid lines from the bottom to the top are density distributions calculated by simulations with magnetic field irregularities for time $t=$ 2,500 to 25,000 $\omega_p^{-1}$ with an interval of 2,500 $\omega_p^{-1}$, respectively. Dotted vertical lines in Figures \ref{dhist6} and \ref{dhist18} show the forbidden distance, which are borders of the forbidden region calculated theoretically in Section \ref{sec:analy} and shown in Figure \ref{penet}.\par

\begin{figure}
	\plotone{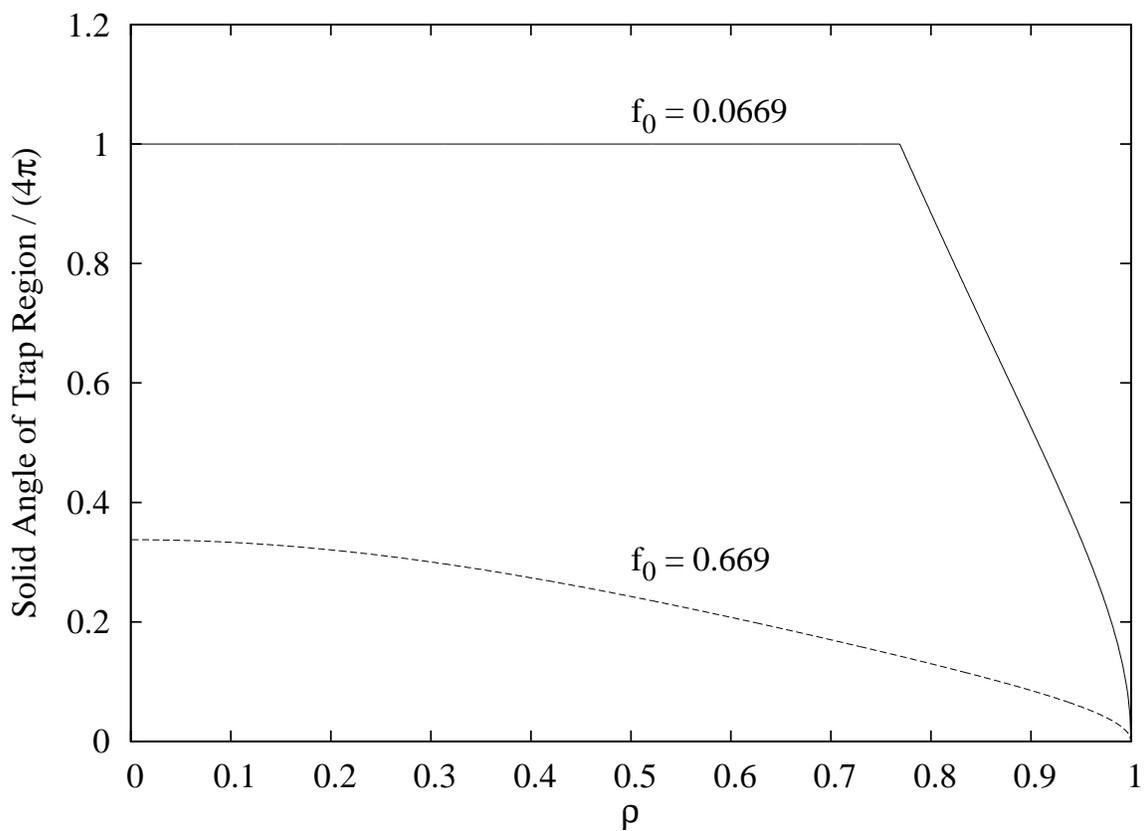}
	\caption{$\rho$ dependence of the solid angle of the trap region normalized by $4\pi$ steradian. Solid and dashed lines are for $f_0=0.0669$ and $f_0=0.669$, respectively. The solid angle of the trap region is larger for smaller $\rho$ for a given $f_0$, and is $4\pi$ steradian in the inner region of the flux rope for $f_0=0.0669$. In this region, all particles are trapped.}
	\label{trap_area}
\end{figure}

We can see from Figure \ref{dhist6}, which shows the case for 6 GV cosmic ray particles, that the density distribution inside and outside the forbidden region is greatly different. Out of the region, there is a density bump around $\rho=0.8$ in the later time. The density out of the forbidden region increases drastically with time and the bump is formed just outside the region. The bump is formed in the following processes. The particles penetrating from the outside of the flux rope edge are scattered by small-scale magnetic field irregularities between the forbidden region and the flux rope edge, and changed their pitch angles. These pitch angle changes can make the particles fall into the trap region, which is expressed as Equation (\ref{Dregion}) and drawn in Figure \ref{trap}. As a result, the particles are trapped between the forbidden region and the flux rope edge. As the solid angle of the trap region near the forbidden region is larger than that near the flux rope edge (see Figure \ref{trap_area}), cosmic ray particles tend to fall in the trap region near the forbidden region, not near the edge. Accumulating these trapped particles, the bump in the density distribution is formed. Moreover, there is a small density drop just inside the flux rope edge. The drop is formed in the following processes. Since the solid angle of trap region is very small just inside the flux rope edge as shown in Figure \ref{trap_area}, nearly particles cannot be trapped in this region. Therefore, the cosmic ray particles in this region are distributed like the steady state distribution without magnetic field irregularities (dashed line in Figure \ref{dhist6}). As the density bump is formed near the forbidden region, this fills in much of the density drop just inside the edge of the flux rope, which is why the small density drop is formed just inside the flux rope edge. A major role of magnetic field irregularities outside of the forbidden region is to make the particles fall into the trap region by pitch angle scattering. Outside of the forbidden region, the effect of a finite Larmor radius is significant for density distribution. Inside of the forbidden region, the trapped particles just out of the forbidden region diffuse into the forbidden region by means of spatial diffusion due to scattering by magnetic field irregularities and a magnetic field line random walk, which is why the density distribution in the forbidden region decreases monotonically toward the flux rope center. Thus, a major role of magnetic field irregularities inside the forbidden region is spatial diffusion toward the flux rope center.\par

Figure \ref{dhist18} shows the results for the case of 18 GV cosmic rays particles. The bump in the density distribution appears while the peak of the bump is located closer to the flux rope axis compared to the case of 6 GV cosmic ray particles. This is because the effect of the Larmor radius for 18 GV cosmic ray particles is greater than that for 6 GV. A density distribution in the forbidden region can be interpreted as the diffusion process as in the case of the 6 GV cosmic ray particles mentioned above.\par

The density distributions for 24 GV and 48 GV cosmic ray particles are shown in Figures \ref{dhist24} and \ref{dhist48}, respectively. These medium- to high-rigidity particles have no forbidden regions. The density distribution for 24 GV cosmic ray particles is almost flat in the inner region although the density peak is formed deep inside in the later time. Although no bump distribution is found in the 48 GV case, the density in the inner region is larger than that in the outer region. Since density in the inner region is smaller than that in the outer region for the density distribution derived from only the diffusion process (Cane \etal 1995), the distributions derived from our simulations, which include an effect of a finite Larmor radius, are reverse distributions compared to those derived from the diffusion process only. In these rigidity ranges, a major role of magnetic field irregularities is to make the particles fall into the trap region by pitch angle scattering. These facts show that the effect of a finite Larmor radius is significant, and we must consider that effect.\par

\section{Discussion}
\label{sec:discussion}
In this study, we used an isotropic magnetic field irregularity model in an interplanetary magnetic flux rope. However, the irregularities in the magnetic flux rope are likely to be anisotropic (Leamon \etal 1998; Narock \& Lepping 2007). Anisotropic magnetic field irregularities cause somewhat large perpendicular diffusion compared to that due to isotropic magnetic field irregularities (Giacalone \& Jokipii 1999). The intensity of magnetic field irregularities ($[\delta B/B]^2=0.01$ in this study) may also be a little different. Therefore, our results may be modified somewhat quantitatively by the anisotropic magnetic field irregularities and its intensity. In our model, we also used a minimum and maximum wave number, and the correlation lengths of the turbulent spectrum were set as the specified values. These values may be a little different from the real situation. We performed many simulations for other parameters and confirmed that they did not cause the results to change qualitatively.\par

Generally, the magnitude of Forbush decreases is highly variable between several to a couple of tens percent for neutron monitor observations (Cane \etal 1996). However, our model shows a much larger decrease (see Figure \ref{dhist6}). The reason for this difference is that simulation time is shorter than propagation time from the Sun to the Earth. The simulation time in our model is 25,000 $\omega_p^{-1}$, corresponding to 0.6 days, because our simulation is time-consuming. In reality, however, the time when a flux rope propagates from the Sun to the Earth is about 1.5 to 2.5 days. If the simulation runs for a longer time, the cosmic ray density in the flux rope will increase and magnitude of decrease will be smaller. We have now investigated more realistic models with longer simulation time.\par

We should discuss the effects of the flux rope curvature and magnetic field configuration for a realistic situation. First, we discuss the effect of curvature of the axis of an interplanetary magnetic flux rope. In the curved flux rope, a cosmic ray particle can have a radial component of gradient-curvature drift velocity and a cosmic ray can penetrate into the flux rope, as discussed by Krittinatham \& Ruffolo (2009). They demonstrated that cosmic ray particles can penetrate into the outer part of a flux rope by means of drift and penetrate into the inner part of a flux rope by diffusion. This is similar to our result, which is the penetration by diffusion into the inner part of the flux rope. On the other hand, the mechanism of penetration into the outer part is different from theirs. This difference is caused by an effect of the finite Larmor radius. Our result shows that cosmic ray particles can penetrate into the outer part of a flux rope via the effect of the finite Larmor radius. Cosmic ray penetration into the curved flux rope with this effect should be further investigated and will advance. Second, we discuss the magnetic field configuration. The roots of magnetic field lines of some interplanetary magnetic flux ropes may be detached from the solar surface and connected to interplanetary magnetic fields (Gosling \etal 1995; Cane \etal 2001). In this situation, cosmic ray particles for a low $f_0$ parameter may be able to penetrate into flux ropes along interplanetary magnetic field lines because the Larmor radius is small compared with the flux rope radius. This may also fill the inner region with cosmic ray particles.\par

An interplanetary magnetic flux rope is often accompanied by a shock wave in front of it. This shock wave can accelerate particles and be one of an origin of solar energetic particles. These energetic particles can also penetrate across the flux rope boundary.  The penetration of these particles is, mainly, diffusion and gradient-curvature drift, and the effect of the finite Larmor radius is small because the particles' energy is several MeV to GeV and the Larmor radius is negligibly small compared to the flux rope radius.\par

\section{Conclusions}
\label{sec:conclusions}
We studied processes for the penetration of cosmic rays into an interplanetary magnetic flux rope, focused on an effect of the finite Larmor radius and magnetic field irregularities. We considered a cylindrical magnetic flux rope as a model of an interplanetary magnetic flux rope. First, we derived an analytical solution for cosmic ray behavior in a magnetic flux rope under the assumption there is no scattering by small-scale magnetic field irregularities. The solution showed that cosmic ray behavior is determined by only the parameter $f_0$; that is, the ratio of the cosmic ray Larmor radius at the flux rope axis to the flux rope radius. We showed that cosmic rays cannot penetrate into the inner region of a flux rope by only gyration and gradient-curvature drift in the case of small $f_0$. Next, we performed a numerical simulation of cosmic ray penetration into an interplanetary magnetic flux rope. The simulation included an effect of small-scale magnetic field irregularities. The results showed that cosmic rays can penetrate into the magnetic flux rope even in the case of small $f_0$ by the effect of small-scale magnetic field irregularities. The result deduced from the simulation is that the high-rigidity cosmic ray penetration is dominated by diffusion when the flux rope is located near the Sun while dominated by gyration when it is located near the Earth. This simulation also showed that a cosmic ray density distribution for the case of moderate to large $f_0$ parameter is greatly different from the distribution deduced from a guiding center approximation shown in previous studies. Our results show that the effects of the finite Larmor radius and magnetic field irregularities cause the bump or reverse cosmic ray density distribution inside the interplanetary magnetic flux rope.\par

The conclusion of this study is that an effect of the finite Larmor radius is significant for the cosmic ray penetration into an interplanetary magnetic flux rope if the Larmor radius of cosmic ray particles in the flux rope is not negligible compared to the flux rope radius.\par

\acknowledgments

We would like to thank the anonymous referee for the useful comments on improving the manuscript.

%%\appendix

\clearpage

\clearpage

\clearpage

\clearpage

\clearpage

\clearpage

\clearpage

\clearpage

\clearpage

\clearpage

\clearpage

\clearpage

\clearpage

\clearpage

\end{document}